\documentclass[conference]{IEEEtran}
\IEEEoverridecommandlockouts
\usepackage{cite}
\usepackage{amsmath,amssymb,amsfonts}
\usepackage{algorithmic}
\usepackage{graphicx}
\usepackage{textcomp}
\usepackage{xcolor}
\usepackage{makecell}
\usepackage{multicol,lipsum}
\usepackage{subcaption}
\usepackage{float}
\usepackage{caption}
\usepackage{caption}
\usepackage[symbol]{footmisc}
\usepackage{fancyhdr}
\usepackage{titlesec}
\usepackage[misc]{ifsym}

\usepackage{url}
\usepackage{verbatim}
\usepackage{filecontents}
\usepackage{pgfplots}
\usepackage{pgfplotstable}
\usepackage{siunitx}    
\usetikzlibrary{pgfplots.statistics}
\pgfplotsset{width=7cm,compat=1.3}
\usetikzlibrary{patterns}
\newcommand{\quotes}[1]{``#1''}
\newcommand{\corrAuthor}{$^{\textrm{(\Letter)}}$}

\def\BibTeX{{\rm B\kern-.05em{\sc i\kern-.025em b}\kern-.08em
    T\kern-.1667em\lower.7ex\hbox{E}\kern-.125emX}}

\begin{document}

\title{A use case of Content Delivery Network raw log file analysis\\
}
\author{\IEEEauthorblockN{Hoang-Loc La\IEEEauthorrefmark{1}\IEEEauthorrefmark{2}\corrAuthor, Anh-Tu Ngoc Tran\IEEEauthorrefmark{1}\IEEEauthorrefmark{2}, Quang-Trai Le\IEEEauthorrefmark{1}\IEEEauthorrefmark{2}, Masato Yoshimi\IEEEauthorrefmark{3}, Takuma Nakajima\IEEEauthorrefmark{4} and Nam Thoai\IEEEauthorrefmark{1}\IEEEauthorrefmark{2}\corrAuthor}
\IEEEauthorblockA{\IEEEauthorrefmark{1}High Performance Computing Laboratory, Advanced Institute of Interdisciplinary Science and Technology, \\ Faculty of Computer Science and Engineering,  Ho Chi Minh City University of Technology (HCMUT), \\268 Ly Thuong Kiet Street, District 10, Ho Chi Minh City, Vietnam \\
\IEEEauthorrefmark{2}Vietnam National University Ho Chi Minh City, \\Linh Trung Ward, Thu Duc District, Ho Chi Minh City, Vietnam\\
Email: \{loclh249, anhtutran, 1652620, namthoai\}@hcmut.edu.vn}
\IEEEauthorblockA{\IEEEauthorrefmark{3}Strategic Technology Center, \IEEEauthorrefmark{4}Service Strategy Sector, \\
TIS Inc., Shinjuku-ku, Tokyo, Japan\\
Email: \{yoshimi.masato, nakajima.takuma\}@tis.co.jp}

}


\maketitle

\begin{abstract}
The growth of video streaming has stretched the Internet to its limitation. In other words, the Internet was originally devised to connect a limited number of computers so that they can share network resources, so the Internet cannot handle a large amount of traffic at a time, which leads to network congestion. To overcome this, CDNs are built on top of the Internet as an overlay to efficiently store and swiftly disseminate contents to users by placing many servers and data centers around the globe. The topic of CDNs has been extensively studied in the last several decades. However, there is still a certain gap between theories in academia and current technologies in industry.
In this paper, we take a close look at the design, implementation, solution, and performance of a CDN system by analyzing its raw log files. Specifically, its infrastructure and system design are first presented, and then we conduct a trace-based study to understand user access patterns, the sources of requests, system performance, and how such information can be used to improve the whole CDN system.
\end{abstract}


\section{Introduction}
\label{Sect:Introduction}
With the growth of user demand on the Internet, there are enormous data that are traversed through the network every day\cite{cisco_forecast}. Essentially, caching techniques are developed, and they permeate almost every area of computer networks. The caching ideas appear from local web browsers\cite{web_caching} to global delivery networks \cite{role_of_caching_in_network}. Content Delivery Networks (CDNs) are one of the large scope caching solutions, which becomes more popular today. CDN providers often use well-known public frameworks to develop their systems\cite{Berger2016AchievingHC} such as Nginx\cite{nignx}, Varnish, and Apache, which makes them do not clearly understand their caching solution. At the same time, the optimal configuration of system hardware and network resources in a dynamic intensity environment is also a critical problem of CDN operators, which can help CDN owners reduce their operating costs and boost up the quality of user experience. 
\par
On the other hand, there is a gap between the theoretical models and the real environment. For instance, Felipe\cite{catalog_dynamics} and Mohamed\cite{temporal_locality} used real logs from telecommunication companies to analyze Video-on-Demand (VoD) data characteristics and from these features, they proposed theoretical models to analyze system performance. Nevertheless, enterprises providing CDN services do not only concern about the hit rate metric, but they also want to know about the quality of service (QoS) and potentials when scaling up their business. In this paper, we adopt a data analysis process and apply it in a real use case. Raw log files of a CDN provider is first filtered and extracted into a meaningful format. After that, we use a visualization tool to get insights into user behavior, content characteristics, and system behavior with dynamic workload intensity. From the extracted information, we can evaluate system performance, quality of user experience, and recommend some ideas to increase enterprises' benefits. 
The remaining sections are organized as follows: The next part reminds background knowledge about the technologies which are applied in the studied system and data processing stages. Section \ref{Sect:infrastructureServices} introduces the role of a CDN system in a general picture,  its detail configurations, and workload types. Section \ref{Sect:traceAnalysis} describes the raw log file format, how we extract useful information from the raw, and our analysis for the extracted information. From these analyses, we also propose ideas to improve system performance and stability. Section \ref{Sect:Related_Work} introduces some related work. Section \ref{Sect:Conclusion} contains some concluding remarks and future work.   
\section{Background}
\label{Sect:Background}
\subsection{Content Delivery Network (CDN)}
Over-The-Top (OTT) services have become more and more popular with a large number of users. The traditional hosting schema, which is only a server or a cluster of servers, cannot satisfy the user demand, especially in case of video streaming. Video streaming services are strictly constrained by delay time, which directly affects user experience. To overcome this, virtually every content provider is currently using CDNs to broadcast their contents\cite{a_tale_3_cdn, neflix_cdn}. CDN is a geo-location distribution network, 
in which each content can have multiple replicas in caching nodes. When a user requests a content, instead of being sent from a far original server to the user's device, the packet can be served from a local caching server.
\subsection{Video Streaming}
The video streaming contributes to major traffic of the OTT service system. Video streaming protocol is a standardized delivery method for breaking up videos into chunks, adapting, and sending them to users. The video chunks will be reassembled at user side. Furthermore, there are several adaptive bitrate protocols to change features of videos in a given time as resolutions and language subtitles. A media streaming process has 3 important stages: encoding, decoding, and transcoding. Encoding is a process compressing large video files for easier uploading to a network. In opposite, the decoding process decompresses the encoded files, expands them to the original form. Moreover, to display video content more flexibly in other devices, platforms, transcoding is applied. Transcoding is a process of taking an encoded stream and modifying some attributes like the size or the encoded bit rate. A transcoder receives an encoded video chunk from an encoder, decodes it into a raw again, changes it in some significant ways, and re-encodes it. For instance, Netflix uses their adaptive streaming algorithm to adjust streaming quality during playback depending on their customers' current network and device condition\cite{netflix_adaptive_stream}.
\par
To distribute adaptively video content for a wide range of users, CDN providers have to transcode source videos into multiple groups, each with different resolutions, data rates. Then they package each group of streams into an adaptive format for multiple users, which includes splitting large videos into smaller chunks and creating a file manifest that indexes the content and their location. Assuming that, they have to provide the content for desktop and mobile devices, each requires a different format. With static packaging, they have to create two separate groups of these packaged assets and upload them to storage servers, which increases the storage cost. Instead of that, they can use dynamic packaging. Regarding dynamic packaging, content providers only need to send a source video to CDN servers. The servers will dynamically transcode the content chunks into multiple formats when the content is requested. There are various streaming protocols as HTTP Live Streaming (HLS)\cite{apple_hls}, Dynamic Adaptive Streaming Over HTTP (DASH)\cite{dash}, Smooth Streaming. In this paper's scope, we only discuss HLS and DASH.
\par
HLS was originally developed by Apple, at first, only the iPhone supported HLS. However, today almost every device supports this protocol. Each HLS segment is usually 10 seconds in duration and has the extension .ts. HLS manifest file, which contains metadata that links to content chunks or other manifest files, has the extension .m3u8. MPEG-DASH is a competitor of HLS, which was created between 2009 and 2012. DASH segments are shorter than HLS, with 2 to 4-second durations being common. Like HLS, DASH also has manifest and chunk files with corresponding extensions: .mpd, .dash. Both protocols are applied in specific content types that depend on the content providers' requirements.

\begin{figure}[H]
\centering
        \includegraphics[totalheight=4.5cm]{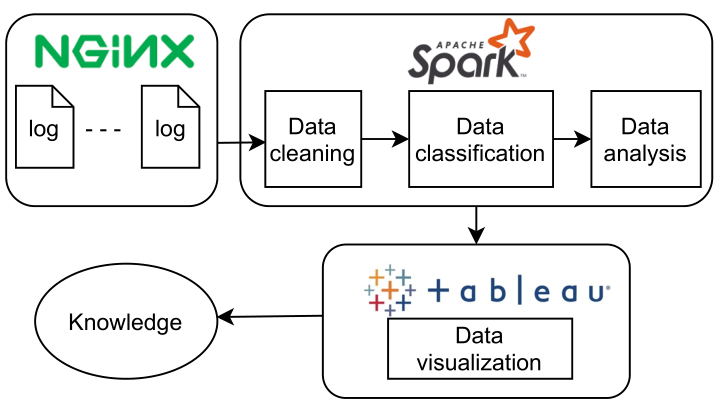}
    \caption{Data processing workflow.}
    \label{fig:data_analysis_structure}
\end{figure}

\subsection{Analysis Framework}
We use Apache Spark\cite{apache_spark} to handle large log files and Tableau\cite{tableau} to visualize the extracted information. Figure \ref{fig:data_analysis_structure} shows an overview of the data processing workflow.
\par

\section{A case study of A Content Delivery Network provider}
\label{Sect:infrastructureServices}
We use real data log files from a large CDN solution vendor in Vietnam which provides the CDN infrastructure and software for content providers. Their customers are pre-dominantly internet television or OTT service suppliers. The CDN infrastructure, which we will consider in this paper, is dedicated to FPT Corporation\cite{fpt_corporation}. FPT is one of the largest ISP companies in Vietnam. They use the CDN system to improve their OTT platform's quality.

\subsection{The CDN provider infrastructure}
\begin{figure}[!htb]
\centering
    \includegraphics[totalheight=6cm]{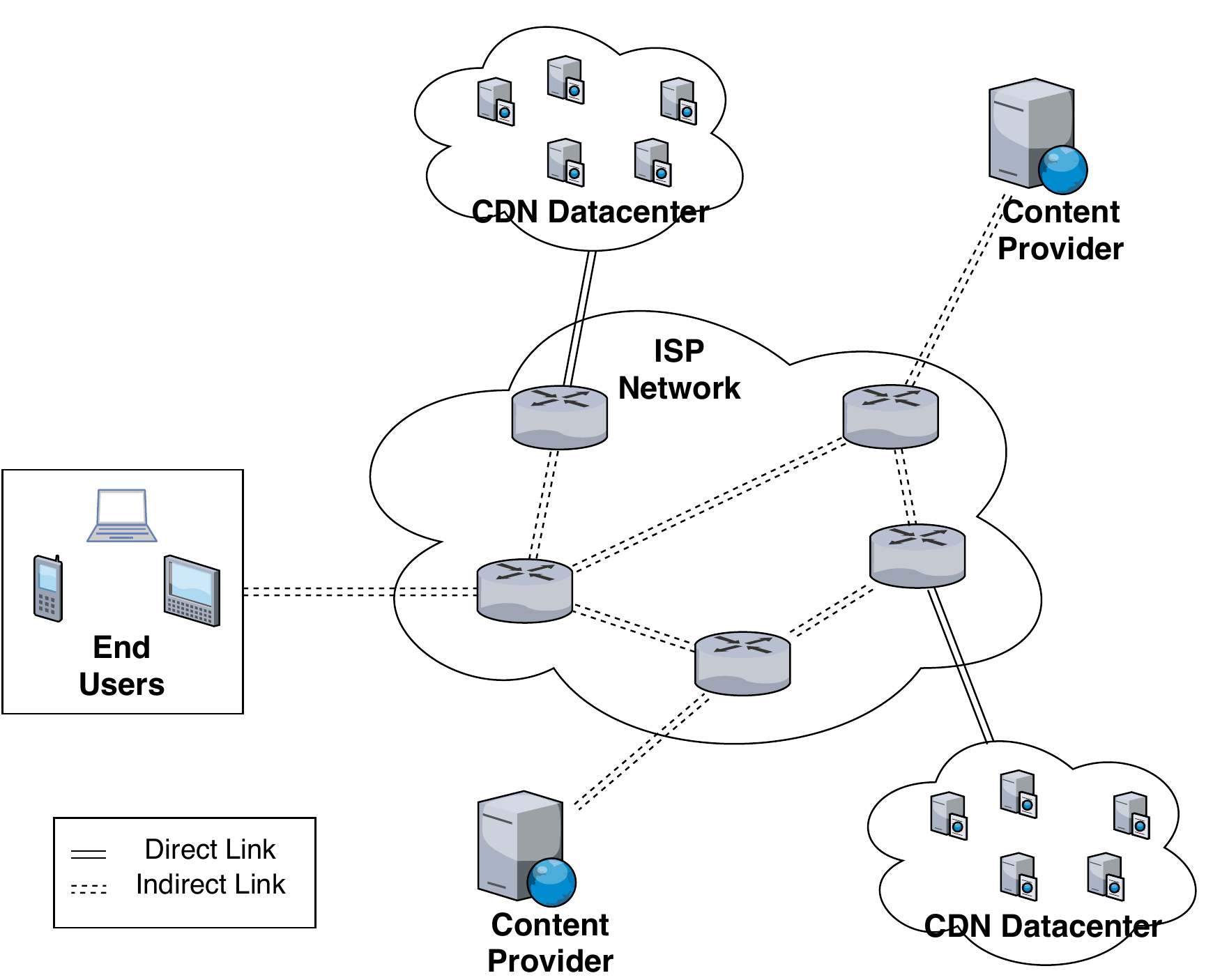}
  \caption{CDN role in ISP network.}
  \label{fig:cdn_isp}
\end{figure}

Depending on the ISP demand, CDN providers will allocate data centers at suitable ISP Point of Presences (PoPs). Figure \ref{fig:cdn_isp} depicts the position of CDN solution in a general ISP network. The content providers and end-users can locate at other ISP networks. The data centers directly connect to ISP PoPs.
\par
\begin{figure}[!htb]
  \centering
    \includegraphics[totalheight=4cm]{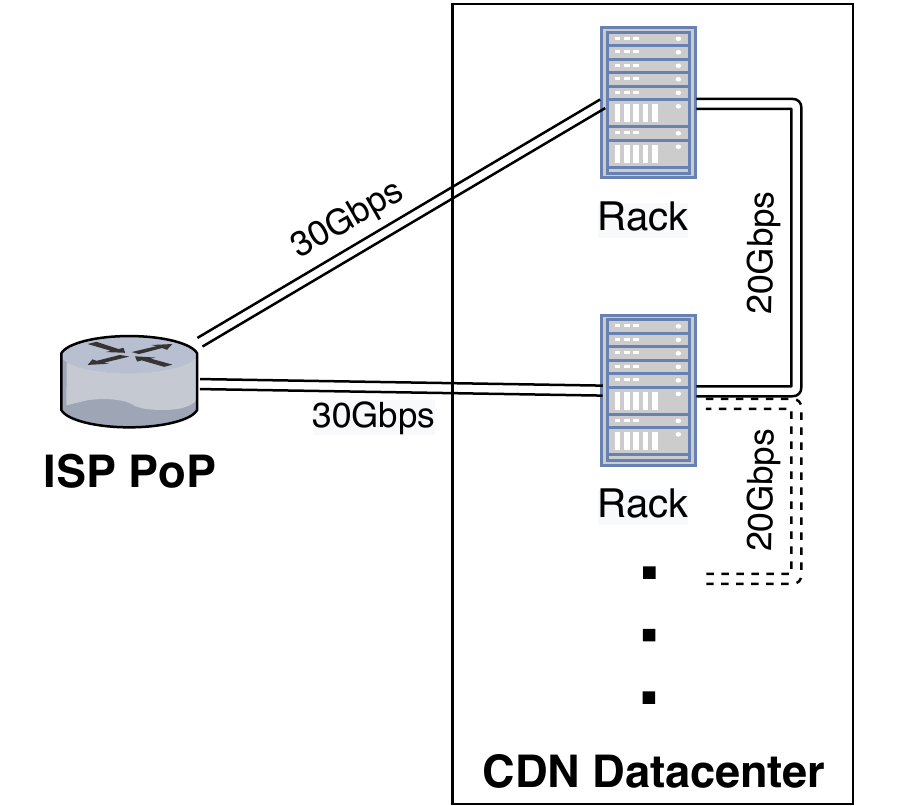}
  \caption{CDN datacenter.}
  \label{fig:cdn_datacenter}
\end{figure}
\begin{figure}[!htb]
  \centering
        \includegraphics[totalheight=4cm]{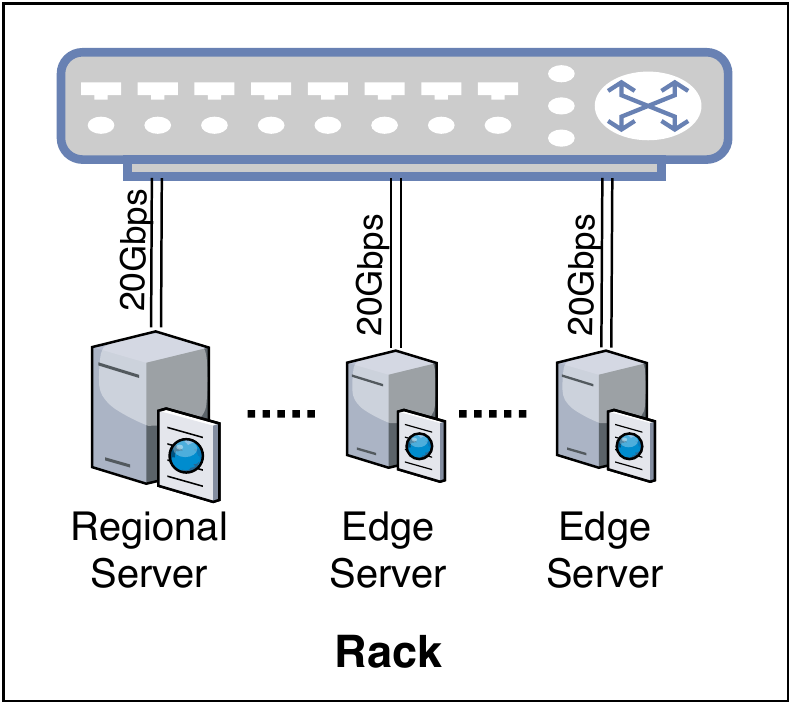}
      \caption{CDN rack.}
      \label{fig:cdn_rack}
  \end{figure}

At each data center, the vendor set up several inter-connected racks. Each rack has a distinct link to an ISP PoP and between each pair of adjacent racks is a connection. In terms of a rack, there are several servers, which connect together with a 48 port switch. More detail for the configuration of the data centers is illustrated in Figures \ref{fig:cdn_datacenter} and \ref{fig:cdn_rack}. 

\begin{figure}[!htb]
\centering
    \includegraphics[totalheight=5cm]{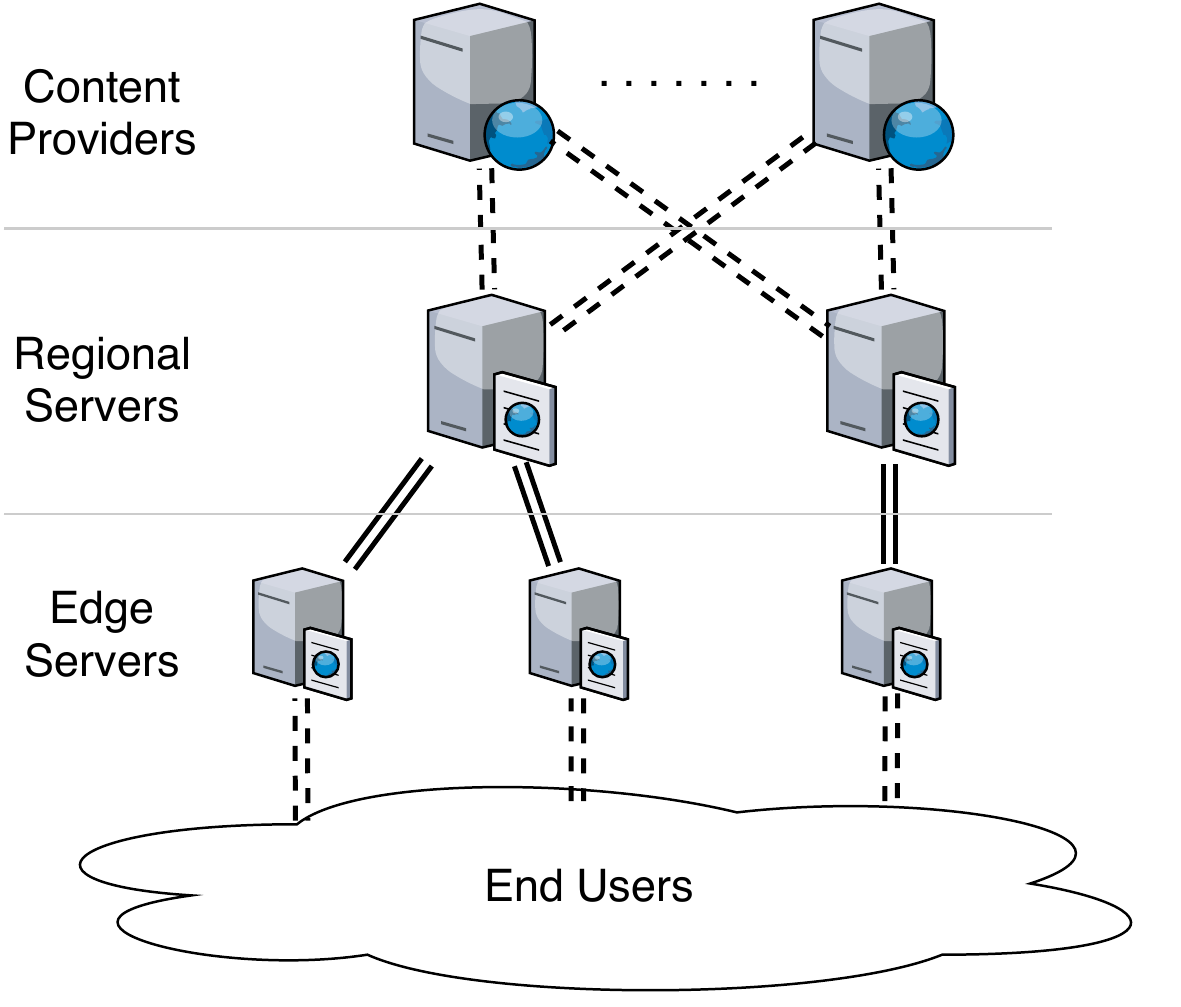}
  \caption{2-layer CDN topology.}
  \label{fig:cdn_2_layer}
\end{figure}

Particularly, the considered provider only allocates one data center and one rack at this data center for FPT. The rack is built with 2 regional servers and 3 edge servers, which constitutes a 2-level hierarchical topology. Each of these servers has about 32GB RAM and about 3TB hard disk. Figure \ref{fig:cdn_2_layer} shows an overview of the system. Least Recently Used (LRU) algorithm is also applied at each cache servers.
When a user requests content, firstly it will be checked in local devices as a web browser and if not, the request will be forwarded to the nearest edge server. If the edge server does not have the content, it will redirect the request to the upper-level server. Again if the requested content is not in the regional cache, it will be sent to content providers, which are also called origin servers. In terms of the responding path of content, the content will be cached at every server in its traversal path. However, packaging contents are pre-stored at regional servers, so the servers will never miss these contents. The details for this content type will be discussed more in the following parts.

\subsection{Workload types} \label{ssec:wordload_types}
\begin{figure}[!htb]
\centering
    \includegraphics[totalheight=6cm]{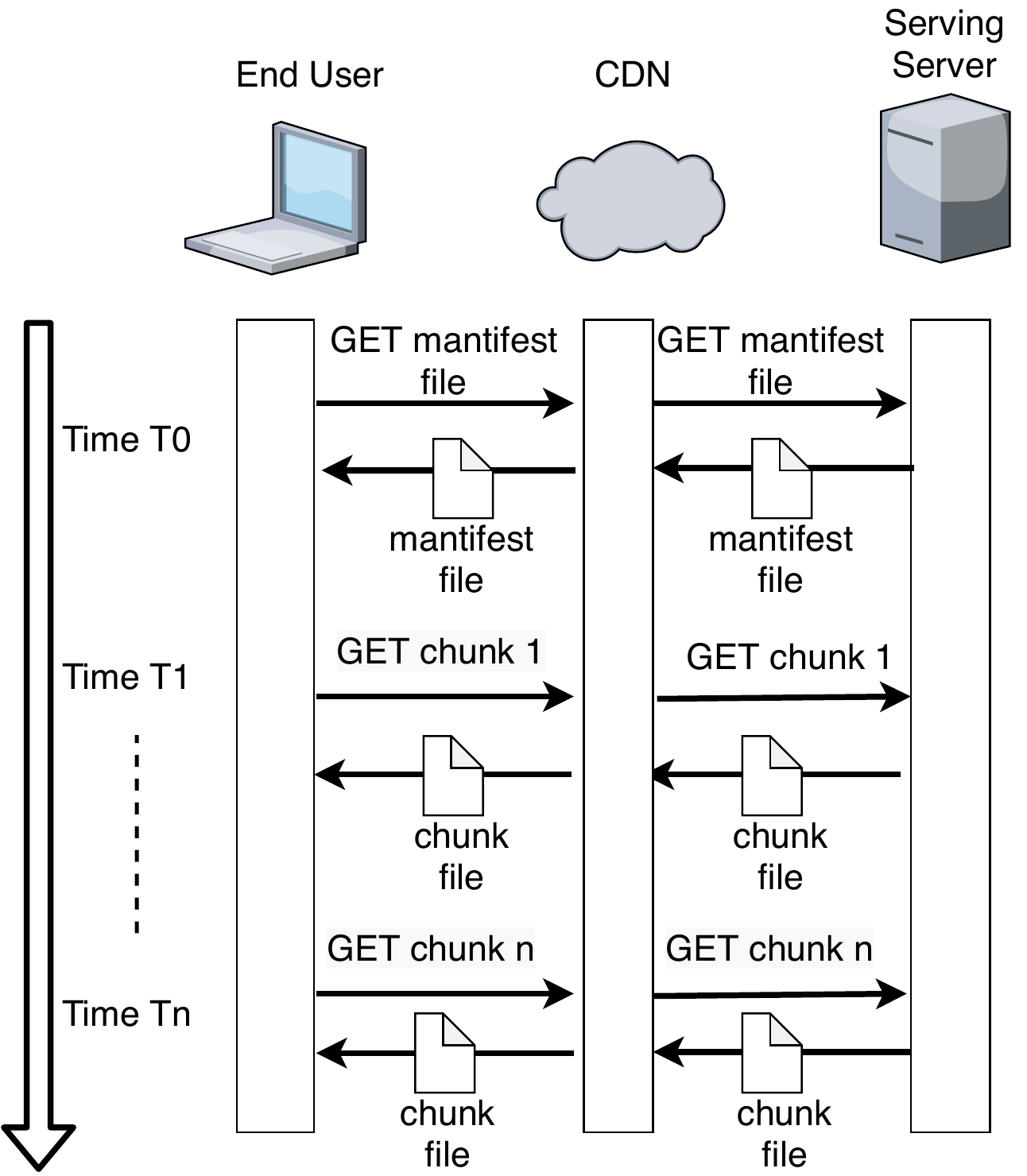}
  \caption{Video streaming workflow.}
  \label{fig:streaming_work_flow}
\end{figure}

The OTT platform mainly supports live streaming, VoD, and web browser services. Streaming videos are applied by both HLS and MPEG-DASH protocols, which are the most popular streaming formats. Depending on the customer demands, the CDN providers use both static content and dynamic packaging mechanism. Figure \ref{fig:streaming_work_flow} depicts the workflow of the video streaming process. In the beginning, the user will request a manifest file, which contains metadata. The original video will be segmented into multiple equal-sized chunks and will be sent to the user when being requested.
\par
The users can watch a video in different platforms or resolutions, which requires the content has to be provided in various formats. There are two ways to solve these problems: prepackaging and statically storing each format copy of original content, or only caching one source content and dynamically packaging it into any format that it needs to be. Clearly, the second approach helps us to reduce the storage cost and be more adaptive for new formats. The CDN vendor applies both the above strategies for their system. In particular, depending on their customer requirements, they will use suitable strategies. With the packing strategy, the content providers pre-send content to regional servers. So content will never miss at regional servers. In regards to non-packaging content, if it is not stored in the upper-level cache, it will be a cache miss at this level and the content request will be sent to the original server.

\section{Trace-based system analysis}
\label{Sect:traceAnalysis}
\subsection{CDN log files}
The CDN provider uses the monitor tools to trace and optimize their system. Nginx is one of the most popular web server frameworks, which also supports monitoring system status. It provides a tracing mechanism and a customizable logging file format with various options.
The CDN vendor uses the logging service of Nginx to monitor the history of user requests, and cache status and network traffic and logs them. We can use these raw log files to analyze the system performance, evaluate the user experience, propose a better solution to reduce the operating cost, and improve the service quality. The enterprise can use this information to maximize their benefit and reduce usage resources.
\begin{table}[!htb]
\begin{center}
    \caption{EXAMPLE OF A RECORD IN SYSTEM LOG}
        \begin{tabular}{ c | c }
           \hline
           0.136 & \makecell{The delay for a content \\ which is requested by a user.} \\ \hline
           118.68.222.40 & \makecell{The IP address of the user who \\requests content from server.} \\ \hline
           MISS & \makecell{The hit status \\of cache servers.} \\ \hline
           [03/Dec/2018:00:00:00, +0700] & \makecell{The time that server receives \\ user request.}  \\ \hline
           \makecell{/38f16b08fdbe06b13a7698f141\\
           672c7a1543774259/tv/\_definst\_/\\dongthap1tv-mid-5803464.ts} & \makecell{The content name which \\the user request.} \\ \hline
           437664 & \makecell{The file size of requested \\content(byte).} \\ \hline
           
        \end{tabular}
        \label{tab:example_log_format}
\end{center}
\end{table}
\vspace{-3mm}
\par Table \ref{tab:example_log_format} shows an example of a raw log file. When a request package goes through the system network, edge servers will monitor their status, and record it. Because their network includes edge cache servers and regional cache servers, hit statuses are denoted as follows:
\begin{itemize}
\item \quotes{MISS}: the content has not been cached at any cache.
\item \quotes{HIT}: the content has been cached at some edge caches.
\item \quotes{HIT1}: the content has been cached at some regional caches.
\item \quotes{-}: the content has been cached at local devices.
\end{itemize}
\subsection{Data analysis}
Because log files of the system can have a very large size (more than 10 GB per day), we use Apache Spark to process and analyze. Raw logs contain noises and error information. Therefore raw data are pre-processed and incorrect records are discarded. The system serves several content providers with multiple service types, where each content type has different characteristics. Consequently log records need to be classified in order to analyze the system more accurately. Then we evaluate the current solution of the CDN provider with hit rate and latency. These metrics will reflect the system status when the workload intensity and content popularity change over time. Besides that, we analyze the data features for each data class. Finally, we use the Tableau tool to visualize the results. 
\subsubsection{Data cleaning}
Data cleaning is an essential stage before we can discover useful knowledge. Each line of the log file relates to a user request record and each field value is split by a comma character. Some of the records do not have enough value of 6 fields as the example in Table \ref{tab:example_wrong_format}, or the value of fields is not correct. Experimental results on our log show that about 0.1\% of records have the wrong format. As shown in Table \ref{tab:example_wrong_format}, the red rows are the discarded records.

\subsubsection{Records Classification}
The log files contain data that belong to several services. Data from each service will have different characteristics that the caching algorithms are very sensitive to. In order to analyze the effect of caching solutions to system performance, we separate the log data into more specific workloads. As discussed in \ref{ssec:wordload_types}, the considered system mainly serves Live Streaming, VoD, and Website services, so we classify the records into these services. Because the content providers have regulated their own content names that relate to the supported service, we can classify the records by their name patterns. 

With live-streaming records, their name contains patterns such as \quotes{live}, \quotes{tv} and the television channel name. The VoD records are the other chunks and manifest file requests, the name of which does not contain any pattern related to live streaming class, will have an extension like: \quotes{.dash}, \quotes{.ts}, \quotes{.m3u8}, \quotes{.mpd}. The remaining records belong to the website class. Moreover, we can split the log into smaller parts to analyze. The vendor uses 2 different packaging mechanisms for dynamic and static content. With dynamic packaging content, the hit value of all requests which asks for these content will be \quotes{HIT}. In the CDN mechanism, a normal content will always have at least one \quotes{MISS}. We can list all the packaging content which their request does not have any \quotes{MISS}. Other content is classified into no-packaging class.
\par
Table \ref{tab:num_rec_per_class} shows the number of records for each class from system logs in 7 days. The number of live streaming requests is more than 94.5\% of the total number of requests for the system.

\subsubsection{Geo-location IP address mapping}
The geo-location of users is important information. The CDN provider can use this information to set up their topology network and allocate cache servers at optimized locations. We use ideas from\cite{nguyen2018analyzing} to build a mapping database. In the first run, the information of the IP address will be requested from an online API\cite{ip_api}. Sometimes these API will return wrong location information. Synonymous location names, which imply the same geo-location but have different names, will be manually grouped. The IPs, which have other wrong location information, will be discarded. This database will be permanently stored in disks. It contains a mapping table between IP addresses with their ISP name and their geo-location. The user geo-location distribution can be easily visualized with Tableau on the world map. 
\begin{table}[!htb]
\centering
        \caption{EXAMPLES OF SOME WRONG RECORDS}
        \begin{tabular}{c}
           \hline
           \makecell[l]{\textcolor{red}{0.017, 118.69.133.153, -, [03/Dec/2018:00:00:00 +0700],} \\ \textcolor{red}{/img\_songs/Nonstop, TONNY}} \\ \hline
           \makecell[l]{0.136, 118.68.222.40, MISS, [03/Dec/2018:00:00:00 +0700], \\ /38f16b08fd/dongthap1tv-mid-5803464.ts,  437664}\\ \hline
           \makecell[l]{\textcolor{red}{0.019, 118.69.133.153, -, [03/Dec/2018:00:00:00 +0700],} \\ \textcolor{red}{/img\_songs/Nonstop, }}
           \\ \hline
           \makecell[l]{0.000, 1.52.122.25, HIT, [03/Dec/2018:00:00:00, +0700], \\ /live/prod\_kplus\_pm\_hd-audio\_vie=56000-video=2499968.m3u8, 0} \\ \hline
        \end{tabular}
        \label{tab:example_wrong_format}
\end{table}
\vspace{-2mm}
\begin{table}[!htb]
\centering
        \caption{NUMBER OF RECORDS PER CLASS}
        \begin{tabular}{|c|c|c|}
           \hline
           \textbf{Service} & \textbf{Is Packaging} & \textbf{Number of Records} \\ \hline
           Live Streaming & No & 152697608 \\ \hline
           Live Streaming & Yes & 129278868  \\ \hline
           Video-On-Demand & No & 2069393 \\ \hline
           Website & No & 14301252 
           \\ \hline
        \end{tabular}
        
        \label{tab:num_rec_per_class}
\end{table} 
\vspace{-3mm}
\subsection{Experiments}
The log files contain transaction records in 7 days, the size of which is approximately 43GB. We will analyze the whole system workload with metrics to get insights about the user request patterns, requested content characteristics, and performance of the current system.
\subsubsection{User Request Distribution and Pattern} 
\begin{figure}[!htb]
\centering
        \includegraphics[totalheight=5.5cm]{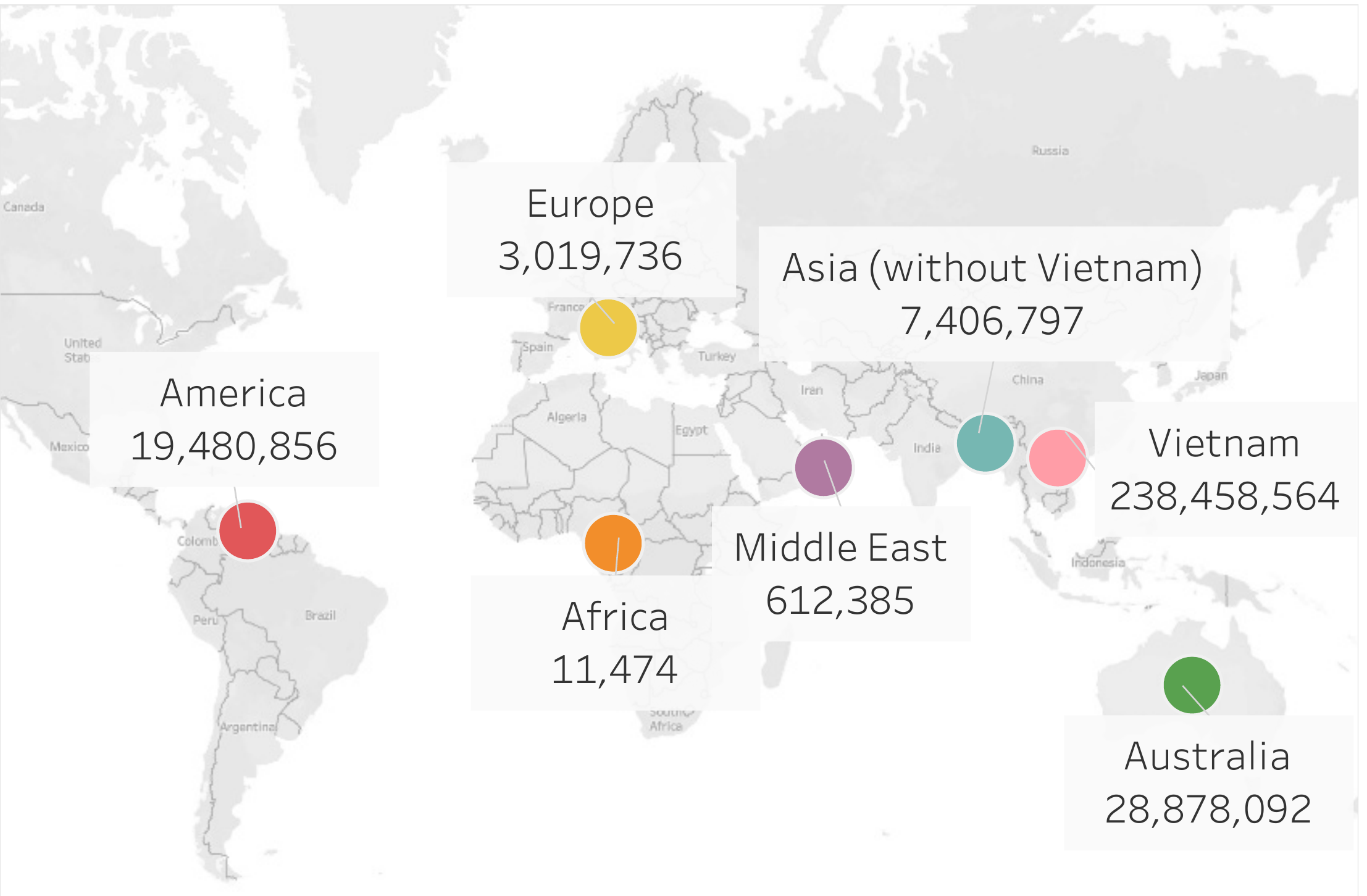}
    \caption{The user request geo-location distribution in world map.}
    \label{fig:geo_location_user_req}
\end{figure}

\begin{figure}[!htb]
\centering
        \includegraphics[totalheight=5.5cm]{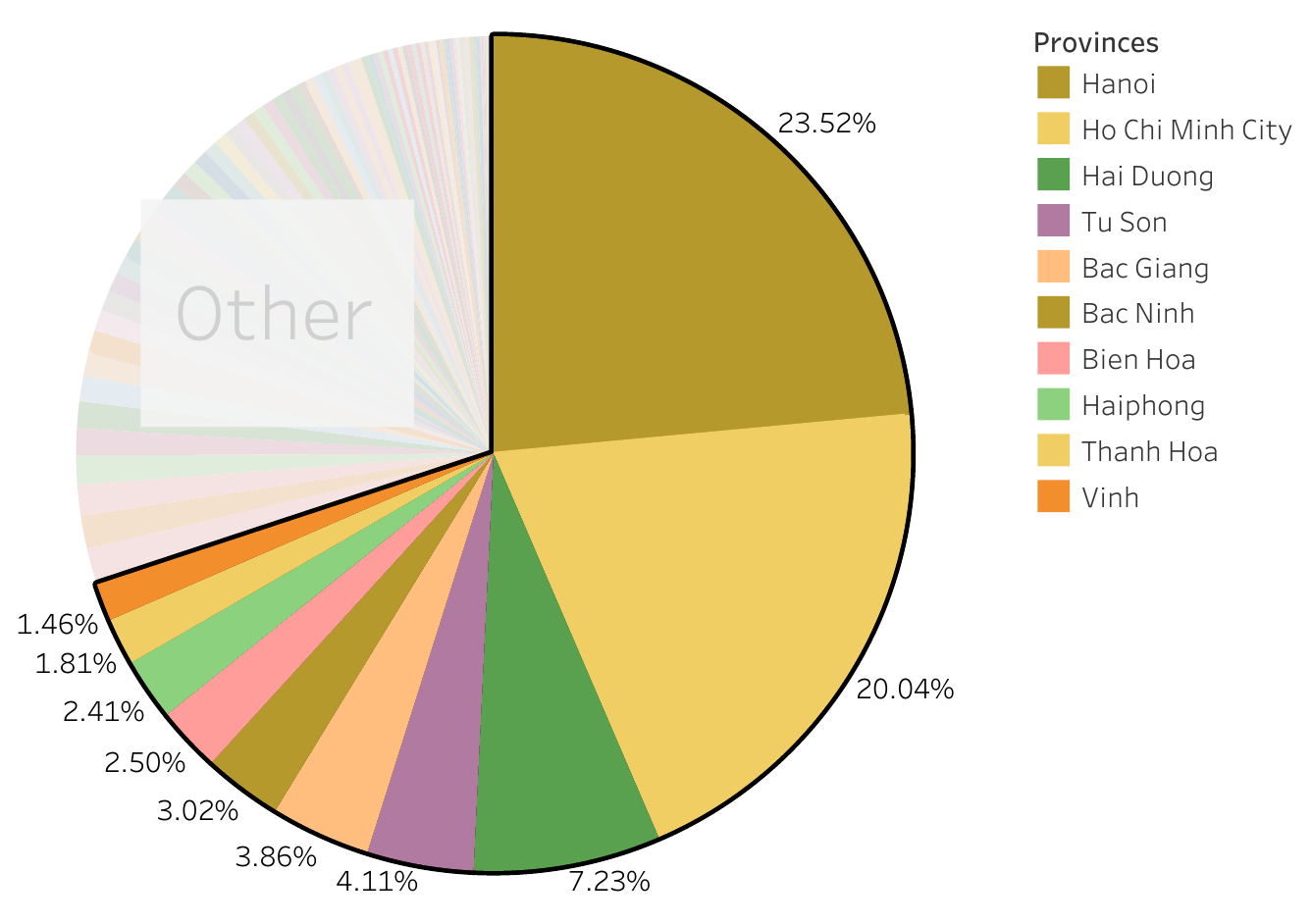}
    \caption{The percentage number of requests of each province in Vietnam.}
    \label{fig:pie_provinces}
\end{figure}

As Figure \ref{fig:geo_location_user_req} shows the geo-location of user requests in 7 days. The system users are not only in Vietnam, but there are a large number of user requests coming from America and Australia. With knowledge of global user distribution, for instance, the CDN provider can add more data centers in appropriate places.
Figure \ref{fig:pie_provinces} illustrates the proportion of request number for provinces in Vietnam. The pie chart shows the percentage of the top 10 provinces, which have more requests than others. The grey part is the percentage of other provinces. Hanoi has the most user request with about 23.52\%, and the Ho Chi Minh city is in the top 2. Almost all provinces in the top 10 belong to Northern Vietnam. With knowledge about the geo-location distribution of user requests, the CDN provider can build a proper system topology, optimize their system configuration as link bandwidth, delay, server capacity or consider setting up more data centers at other ISPs PoPs.

\begin{figure}[!htb]
\centering
    \includegraphics[totalheight=3.5cm]{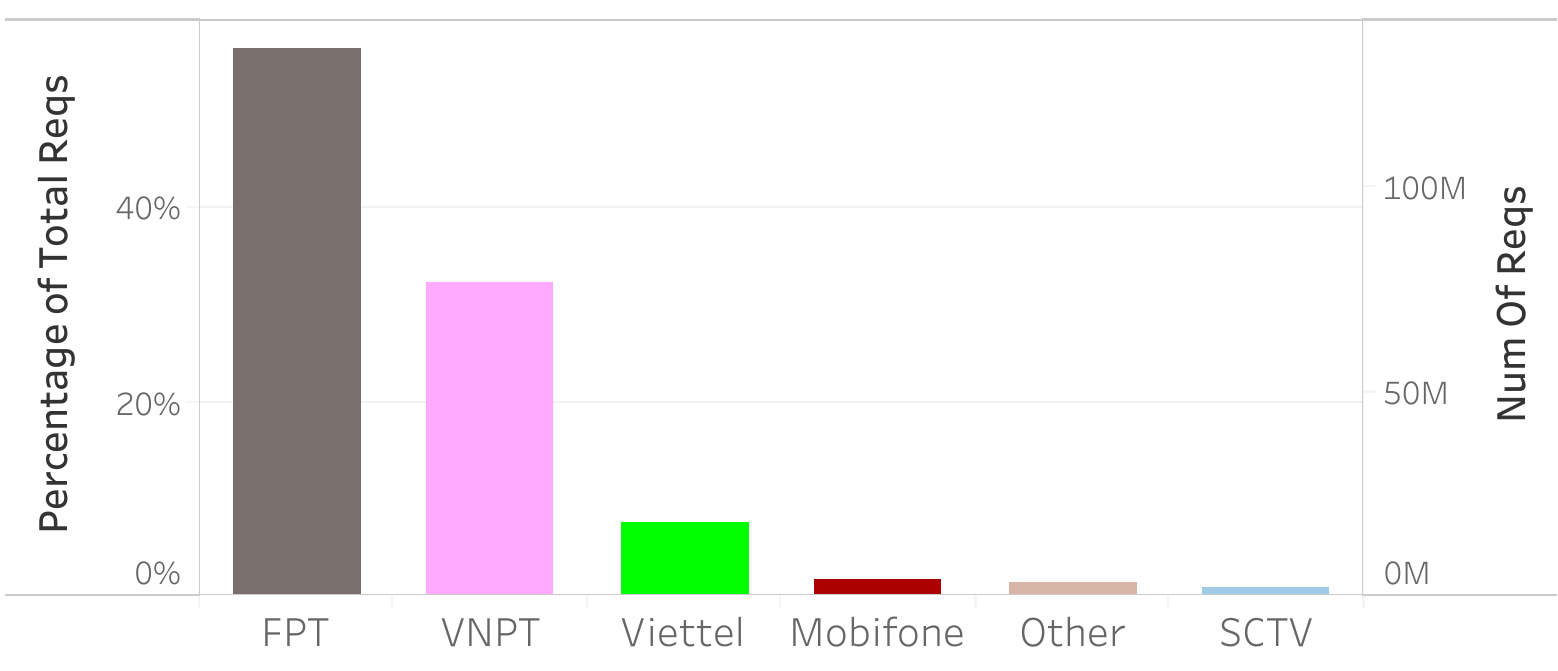}
    \caption{The proportion of number of requests between ISPs.}
    \label{fig:isp_count_reqs}
\end{figure}

The information about the ISPs of user IPs is necessary for the system manager. Figure \ref{fig:isp_count_reqs} depicts the number of requests from different ISPs. The considered system in this paper serves FPT, that is why their user request takes the largest percentage of the total. Moreover, VNPT and Viettel also contribute large parts of the total. In Section \ref{ssec:overal_performance}, we will focus on these ISPs. 

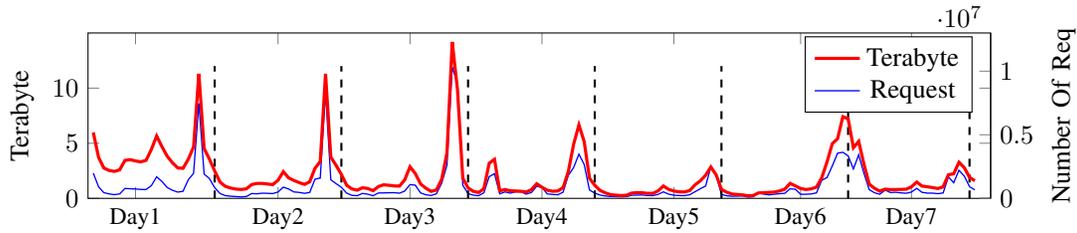
\begin{figure*}
\centering
     \begin{tikzpicture}
        \pgfplotstableread{traffic_data.csv}{\Traffic}
        \pgfplotstableread{request_data.csv}{\Request}
        \begin{axis}[
             scale only axis,
             ylabel = Terabyte,
             height =2.2cm,
             width = 12cm,
             ymin = 0, ymax = 15,
             xmin = -1, xmax = 170,
             ytick = {0, 5, 10},
             xticklabel style = {font=\small},
             xticklabels from table = {\Traffic}{Day}, xtick = {8, 35, 60, 85, 110, 134, 155},
            ]
            \addplot[line width=0.5pt, blue] table [x = i, y = Terabyte] {\Traffic};
            \label{plot_one};
            \draw[black, dashed, thick] (24, 0) -- (24, 120); 
            \draw[black, dashed, thick] (48, 0) -- (48, 120);
            \draw[black, dashed, thick] (72, 0) -- (72, 120);
            \draw[black, dashed, thick] (96, 0) -- (96, 120);
            \draw[black, dashed, thick] (120, 0) -- (120, 120);
            \draw[black, dashed, thick] (144, 0) -- (144, 120);
            \draw[black, dashed, thick] (167, 0) -- (167, 120);
          \end{axis}
          \begin{axis}[
            scale only axis,
            axis y line*=right,
            axis x line=none,
            height = 2.2cm,
            width =  12cm,
            xmin = -1, xmax = 170,
            ymin = 0, ymax = 13000000,
            ytick = {0, 5000000, 10000000},
            ylabel = Number Of Req,
            ]
            \addplot[very thick, red] table [x = i, y = Request] {\Request};
            \addlegendimage{/pgfplots/refstyle=plot_one}\addlegendentry{Terabyte};
            \addlegendentry{ Request};
          \end{axis}
    \end{tikzpicture}
    \caption{The number of user request and total sent file size in 7 days.}
    \label{fig:num_req_7_day}
\end{figure*}

Figure \ref{fig:num_req_7_day} shows the user access pattern in 7 days. The number of requests has a common pattern, which is the highest at 7-9 PM and the second highest at 3-5 AM every day. This pattern is reasonable with the usage habits of users. In Vietnam, people usually use the television in the evening and watch soccer matches in the early morning. The total sent file size has a similar pattern with the number of requests. With this information, the system administrator can dynamically allocate more resources at peak time, and deallocate resources at an off-peak time to reduce the operating cost.

\subsubsection{Content Characteristics} 
\begin{figure}[!htb]
\centering
    \begin{tikzpicture}
        \begin{axis}[
            xbar=0pt, xmin=0,
            width=8cm, height=6.8cm,
            xmax = 1,
            symbolic y coords={m3u8, mpd, ts,  dash, mp3, mp4, image, other},
            ytick=data,
            xmajorticks=false,
            nodes near coords, every node near coord/.append style={font=\scriptsize},nodes near coords align={horizontal},
        ]
           \addplot  [fill=blue, point meta={x*100}, bar width=8pt,
            nodes near coords={\pgfmathprintnumber\pgfplotspointmeta\%}, postaction={pattern=vertical lines}] coordinates {(0.31048,m3u8) (0.08053,mpd) (0.37198,ts) (0.18326,dash) (0.00011,mp3) (0.0027,mp4) (0.04666,image) (0.0067,other)} ;
            \addplot [fill=pink, bar width=8pt, nodes near coords, postaction={pattern=dots}] coordinates {(0.00094,m3u8) (0.0021,mpd) (0.82127,ts) (0.17247,dash) (0.00105,mp3) (0.00186,mp4) (0.00018,image) (0.00013,other)} ;
            \legend{Requests, Bytes};
        \end{axis}
    \end{tikzpicture}
    \caption{The proportion of number of requests and content size for Multipurpose Internet Mail Extensions (MIME) types.}
    \label{fig:file_type_req_size}
\end{figure}
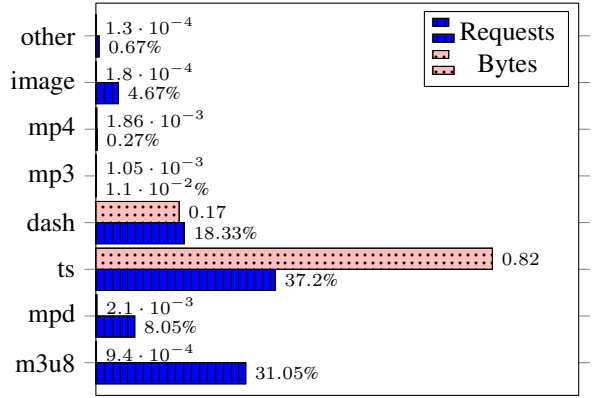

Understanding the requested data features helps the enterprise choose an optimal caching solution. Figure \ref{fig:file_type_req_size} shows that the system mainly handles manifest (mpd and m3u8 files) and chunk files (ts and dash files), which are used in VoD and live streaming services. 

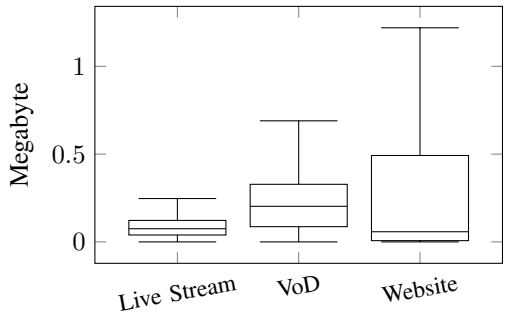
\begin{figure}[!htb]
\centering
    \begin{tikzpicture}
        \begin{axis}[
        boxplot/draw direction=y,
        xticklabel style = {rotate = 10},
        width=7cm, height=5cm,
        ylabel = Megabyte,
        xtick ={1, 2, 3},
        xticklabel style = {font=\small},
        xticklabels = {Live Stream, VoD, Website},
        ]
        \addplot[black, boxplot prepared={ 
            lower whisker=0.00000001, lower quartile=0.03936665, 
            median=0.075, upper quartile=0.12234965,
            upper whisker=0.2468137}]
            coordinates {};
        \addplot[black, boxplot prepared={ 
            lower whisker=0.00000001, lower quartile=0.0866492, 
            median=0.2026076, upper quartile=0.3279754,
            upper whisker=0.6894712}]
            coordinates {};
         \addplot[black, boxplot prepared={ 
            lower whisker=0.00000001, lower quartile=0.00713175, 
            median=0.0578163 , upper quartile=0.49207115,
            upper whisker=1.2192666}]
            coordinates {};
        \end{axis}
    \end{tikzpicture}
    \caption{The file size distribution for each service.}
    \label{fig:file_size_distribution}
\end{figure}

File size distribution is shown in Figure \ref{fig:file_size_distribution}. The live streaming service has a huge number of requests and requested chunks of each specific video are equally segmented, and the segment length is usually shorter than VoD’s segment length. Therefore, its file size variance is smallest. In contrast, the VoD class has the least number of user requests, and its chunks are longer than live streaming's chunks, so its requested size variance is larger. The website service file size is more various. It includes several file types, for instance, file downloading requests, each of which requires different sizes. Although some downloading files can have a huge size, the main part of website files are front end scripts, which have very small sizes, so the median of website file size is the smallest.

\subsubsection{Overall Performance} \label{ssec:overal_performance}
\begin{figure*}
\centering
    \begin{subfigure}[b]{\textwidth}
        \centering
        \begin{tikzpicture}
     \pgfplotstableread{regional_data.csv}{\Regional}
     \pgfplotstableread{edge_data.csv}{\Edge}
     \pgfplotstableread{system_data.csv}{\System}
     \begin{axis}[
         ylabel = Hit Rate,
         height =3.7cm,
         width = 16cm,
         ymin = 0, ymax = 1,
         xmin = -1, xmax = 170,
         ytick = {0, 0.5 ,1},
         xticklabel style = {font=\small},
         xticklabels from table = {\Regional}{Day}, xtick = {8, 35, 60, 85, 110, 134, 155},
         legend style={at={(0.5, 1)},
         anchor=south,legend columns=-1}
        ]
        \addplot[very thick, red] table [x = i, y = p] {\Regional};
        \draw[black, dashed, thick] (24, 0) -- (24, 120); 
        \draw[black, dashed, thick] (48, 0) -- (48, 120);
        \draw[black, dashed, thick] (72, 0) -- (72, 120);
        \draw[black, dashed, thick] (96, 0) -- (96, 120);
        \draw[black, dashed, thick] (120, 0) -- (120, 120);
        \draw[black, dashed, thick] (144, 0) -- (144, 120);
        \draw[black, dashed, thick] (167, 0) -- (167, 120);
        \addplot[very thick, orange] table [x = i, y = p] {\Edge};
        \addplot[very thick, blue] table [x = i, y = p] {\System};
        \legend{Regional Hit Rate, Edge Hit Rate, System Hit Rate}
      \end{axis}
    \end{tikzpicture}

        \includegraphics[totalheight=2.6cm]{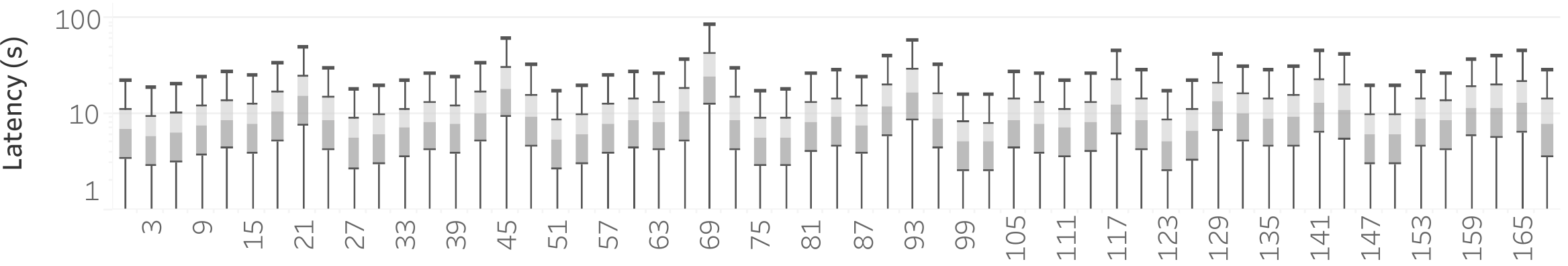}
    \end{subfigure}
    \caption{Upper: General hit rate for 3 services, Lower: Sum latency for 3 services.}
    \label{fig:overall_performance}
    
\end{figure*}

In this part, we will evaluate the latency and hit rate of the system for 3 services. The system hit rate reflects the caching solution's effectiveness. Latency directly impacts the quality of user experience.
In the upper of Figure \ref{fig:overall_performance}, the hit rate pattern of the overall system and edge servers are similar to user request patterns. The edge hit rate and whole hit rate are approximately the same and much larger than the regional hit rate, which means that almost the content is cached by edge servers. In other words, the number of "HIT" records is far greater than "HIT1" records. The other reason is the regional caches adapt to dynamic content popularity slower than edge cache\cite{hierachial_caching}. When the number of user requests increases, the hit rate also rises. So as to evaluate the system adaptive capacity when the number of income requests increases, the bottom of Figure \ref{fig:overall_performance} depicts the latency of the system in 7 days. In peak intervals, the average latency also reaches summits and the upper hinge of latency reaches more than 40s, which can cause long stalls for 25\% of user requests.
\begin{figure}[!htb]
\centering
    \begin{tikzpicture}
        \begin{axis}[
            /pgf/number format/1000 sep={},
            width=6.5cm,
            height=2cm,
            at={(0.758in,0.981in)},
            scale only axis,
            clip=false,
            separate axis lines,
            axis on top,
            xmin=0.5,
            xmax=6.5,
            xtick={1,2,3,4, 5, 6},
            x tick style={draw=none},
            xticklabels={FPT, VNPT, Viettel, Mobifone, Other, SCTV},
            xticklabel style = {rotate = 15},
            xticklabel style = {font=\small},
            ytick={0, 0.2, 0.4,  0.6, 0.8, 1},
            ymin=0,
            ymax=1,
            ylabel={Hit Rate},
            every axis plot/.append style={
              ybar,
              bar width=0.5cm,
              bar shift=0pt,
              fill
            }
          ]
          \addplot[blue]coordinates {(1, 0.8570)};
          \addplot[green]coordinates{(2, 0.8666)};
          \addplot[purple]coordinates{(3, 0.8569 )};
          \addplot[orange]coordinates{(4, 0.9395)};
          \addplot[black]coordinates{(5, 0.9706)};
          \addplot[yellow]coordinates{(6, 0.5811)};
        \end{axis}
      \end{tikzpicture}
    \caption{Hit rate for each ISP.}
    \label{fig:isp_hit_rate}
\end{figure}
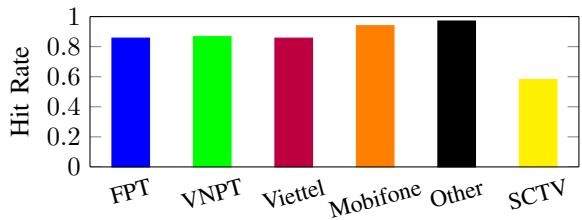

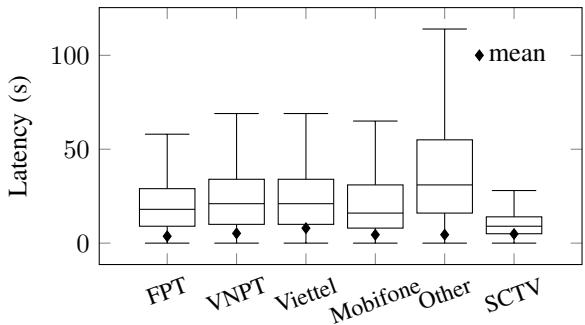
\begin{figure}[!htb]
\centering
    \begin{tikzpicture}
        \begin{axis}[
        boxplot/draw direction=y,
        width=8cm, height=5cm,
        ylabel = Latency (s),
        bar width = 0.5cm,
        xtick ={1, 2, 3, 4, 5, 6, 7},
        xticklabels = {FPT, VNPT, Viettel, Mobifone, Other, SCTV},
        xticklabel style = {rotate = 20},
        xticklabel style = {font=\small},
        ]
        
        \addplot[black, boxplot prepared={ 
            lower whisker=0, lower quartile=9, 
            median=18, upper quartile=29, average=3.67,
            upper whisker=58}]
            coordinates {};
        \addplot[black, boxplot prepared={ 
            lower whisker=0, lower quartile=10, 
            median=21, upper quartile=34, average=5.195,
            upper whisker=69}]
            coordinates {};
         \addplot[black, boxplot prepared={ 
            lower whisker=0, lower quartile=10, 
            median=21, upper quartile=34, average=7.965,
            upper whisker=69}]
            coordinates {};
        \addplot[black, boxplot prepared={ 
            lower whisker=0, lower quartile=8, 
            median=16, upper quartile=31, average=4.462,
            upper whisker=65}]
            coordinates {};
        \addplot[black, boxplot prepared={ 
            lower whisker=0, lower quartile=16, 
            median=31, upper quartile=55, average=4.499,
            upper whisker=114}]
            coordinates {};
        \addplot[black, boxplot prepared={ 
            lower whisker=0, lower quartile=5, 
            median=9, upper quartile=14, average=4.880,
            upper whisker=28}]
            coordinates {};
        \addplot [ mark=diamond*, nodes near coords=mean,every node near coord/.style={anchor=180}] coordinates {( 5.5, 100)};
        \end{axis}
    \end{tikzpicture}
    \caption{Latency for each ISP.}
    \label{fig:isp_latency}
\end{figure}

In addition, we evaluate the influences of the caching solution to users from different ISPs. Figures \ref{fig:isp_hit_rate} and \ref{fig:isp_latency} show the performance metrics of the top 5 Vietnam ISPs. From Figure \ref{fig:isp_latency}, we can conclude that the latency distribution of each ISP has a positive skewness. The considered CDN system mainly serves FPT platform, and we expect that the system will be optimal for FPT users. The hit rate of top 3 ISPs, which contribute the majority of traffic, are approximately the same. Besides, average latency of FPT is also better than the remaining ISPs. Nevertheless, if their workload intensity increases, their solution may not serve  FPT users well. The CDN owner should consider adapting a suitable mechanism to filter requests of other ISP users.

\subsubsection{Performance of Different Workloads}
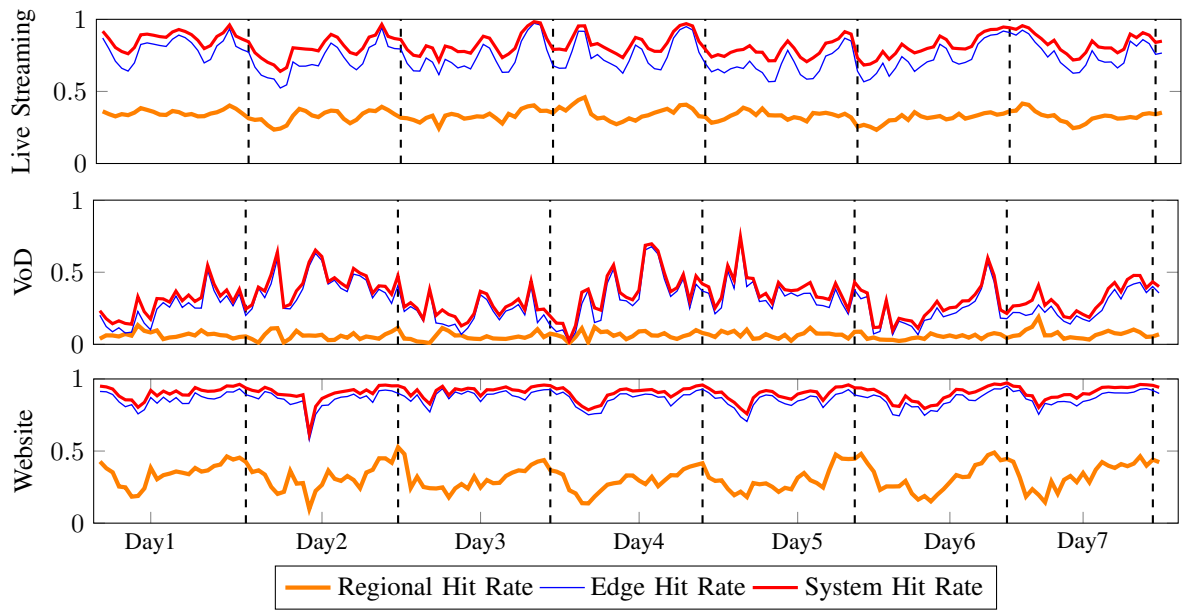
\begin{figure*}
\centering
    \begin{tikzpicture}
     \pgfplotstableread{livestream_regional_data.csv}{\Regional}
     \pgfplotstableread{livestream_edge_data.csv}{\Edge}
     \pgfplotstableread{livestream_system_data.csv}{\System}
     \begin{axis}[
         ylabel = Live Streaming,
         height = 3.5cm,
         width = 16cm,
         ymin = 0, ymax = 1,
         xmin = -1, xmax = 170,
         ytick = {0, 0.5 , 1},
         xmajorticks=false,
         legend style={at={(0.5, -1.1333)},
         anchor=south,legend columns=-1}
        ]
        \addplot[line width=1.6pt, orange] table [x = i, y = p] {\Regional};
        \draw[black, dashed, thick] (24, 0) -- (24, 120); 
        \draw[black, dashed, thick] (48, 0) -- (48, 120);
        \draw[black, dashed, thick] (72, 0) -- (72, 120);
        \draw[black, dashed, thick] (96, 0) -- (96, 120);
        \draw[black, dashed, thick] (120, 0) -- (120, 120);
        \draw[black, dashed, thick] (144, 0) -- (144, 120);
        \draw[black, dashed, thick] (167, 0) -- (167, 120);
        \addplot[line width=0.5pt, blue] table [x = i, y = p] {\Edge};
        \addplot[very thick, red] table [x = i, y = p] {\System};
      \end{axis}
    \end{tikzpicture}
    \begin{tikzpicture}
     \pgfplotstableread{video_regional_data.csv}{\Regional}
     \pgfplotstableread{video_edge_data.csv}{\Edge}
     \pgfplotstableread{video_system_data.csv}{\System}
     \begin{axis}[
         ylabel = VoD,
         height = 3.5cm,
         width = 16cm,
         ymin = 0, ymax = 1,
         xmin = -1, xmax = 170,
         ytick = {0, 0.5 ,1},
         xmajorticks=false,
         legend style={at={(0.5, -1.1333)},
         anchor=south,legend columns=-1}
        ]
        \addplot[line width=1.6pt, orange] table [x = i, y = p] {\Regional};
        \draw[black, dashed, thick] (24, 0) -- (24, 120); 
        \draw[black, dashed, thick] (48, 0) -- (48, 120);
        \draw[black, dashed, thick] (72, 0) -- (72, 120);
        \draw[black, dashed, thick] (96, 0) -- (96, 120);
        \draw[black, dashed, thick] (120, 0) -- (120, 120);
        \draw[black, dashed, thick] (144, 0) -- (144, 120);
        \draw[black, dashed, thick] (167, 0) -- (167, 120);
        \addplot[line width=0.5pt, blue] table [x = i, y = p] {\Edge};
        \addplot[very thick, red] table [x = i, y = p] {\System};
      \end{axis}
    \end{tikzpicture}
    \begin{tikzpicture}
     \pgfplotstableread{website_regional_data.csv}{\Regional}
     \pgfplotstableread{website_edge_data.csv}{\Edge}
     \pgfplotstableread{website_system_data.csv}{\System}
     \begin{axis}[
         ylabel = Website,
         height = 3.5cm,
         width = 16cm,
         ymin = 0, ymax = 1,
         xmin = -1, xmax = 170,
         ytick = {0, 0.5 ,1},
         xticklabel style = {font=\small},
         xticklabels from table = {\Regional}{Day}, xtick = {8, 35, 60, 85, 110, 134, 155},
         legend style={at={(0.5, -0.6)},
         anchor=south,legend columns=-1}
        ]
        \addplot[line width=1.6pt, orange] table [x = i, y = p] {\Regional};
        \draw[black, dashed, thick] (24, 0) -- (24, 120); 
        \draw[black, dashed, thick] (48, 0) -- (48, 120);
        \draw[black, dashed, thick] (72, 0) -- (72, 120);
        \draw[black, dashed, thick] (96, 0) -- (96, 120);
        \draw[black, dashed, thick] (120, 0) -- (120, 120);
        \draw[black, dashed, thick] (144, 0) -- (144, 120);
        \draw[black, dashed, thick] (167, 0) -- (167, 120);
        \addplot[line width=0.5pt, blue] table [x = i, y = p] {\Edge};
        \addplot[very thick, red] table [x = i, y = p] {\System};
        \legend{Regional Hit Rate, Edge Hit Rate, System Hit Rate}
      \end{axis}
    \end{tikzpicture}
    \caption{Hit rate of each service.}
    \label{fig:specific_hit_rate}
\end{figure*}

\begin{figure*}
    \centering
    \includegraphics[totalheight=6.62cm]{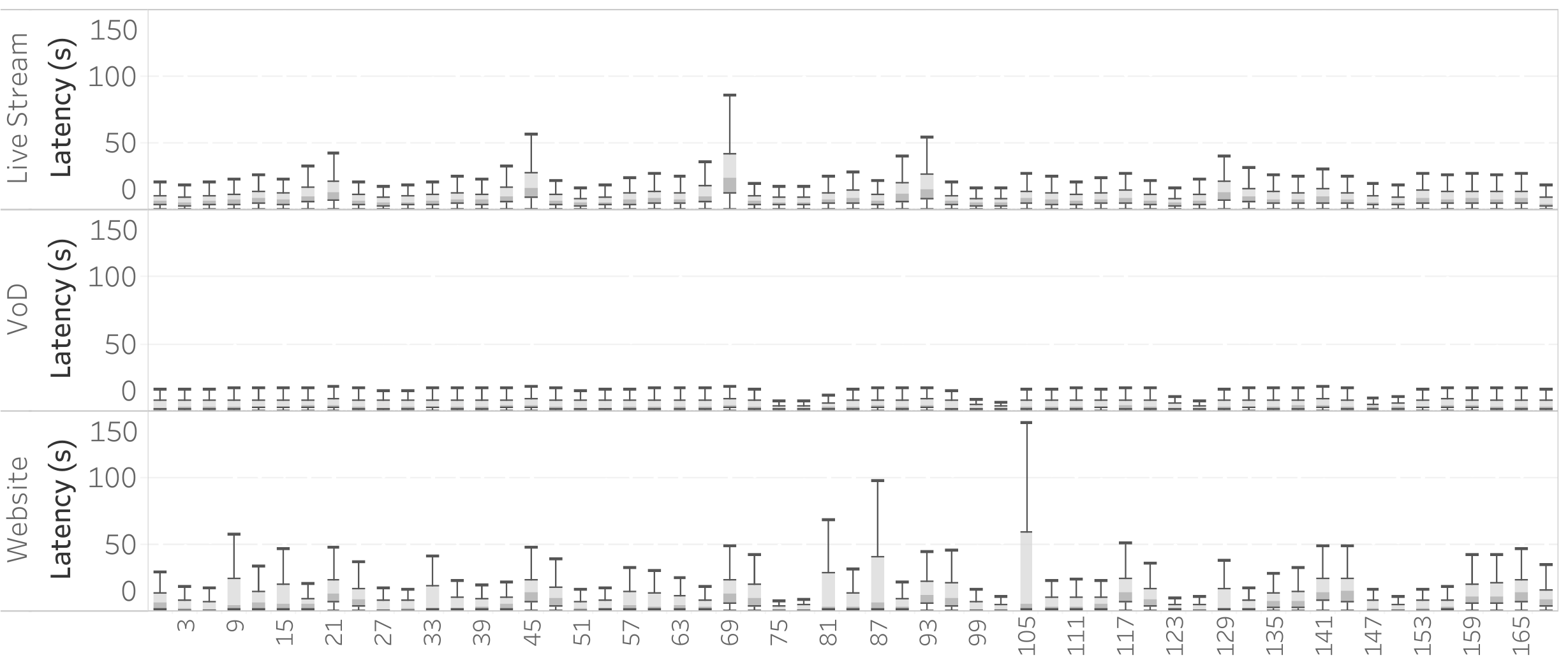}
    \caption{Latency of each service.}
    \label{fig:specific_latency}
\end{figure*}

Figures \ref{fig:specific_hit_rate} and \ref{fig:specific_latency} show the performance metrics of each workload. The live streaming service hit rate lines are very close to the overall service hit rate lines in the upper part of Figure \ref{fig:overall_performance} because the live streaming service dominates the traffic of the system. Although VoD contents are stored on hard disks, they have their own Time-to-Lives (TTLs). When the arrival-time of two consecutive requests of the same content exceeds the content's TTL, the latter request will be \quotes{MISS}. The VoD service's hit rate implies that video content is requested at a very low frequency. The characteristic of VoD contents also explains their steady request latencies. The live streaming contents are cached into RAM with limited size and they contribute the majority of network traffic. The pattern of their latency variance changes similar to user request pattern and reaches high variances at some peaks, which means it still has issues in their caching solution. Although the website workload just contributes the minority of traffic, arbitrariness of the downloaded file sizes creates the diversity of the latencies.

\vspace{-8mm}



\section{Related Work}
\label{Sect:Related_Work}
The topic of CDNs has been extensively studied in the last several decades. However, there is still a certain gap between theories in academia and current technologies in industry. Zolfaghari et al. \cite{zolfaghari2020content} surveys state of the art technologies, academic studies, and current and emerging trends in CDNs. 
To the best of our knowledge, this is the first work taking a close look at CDN log files to analyze in details the system performance of a CDN provider. Information from the CDN log files not only can give us the status of the system but also insights into how to optimize the current system. Furthermore, we present the infrastructure of a CDN vendor in Vietnam. Besides, our work provides some performance criteria that can be considered as a use case for system analysis.

On the contrary, there are still several studies that are moderately relevant to our work. Balachandran et al. \cite{balachandran2013analyzing} observe several user access patterns, regional interests and how such information have important implications to two trending CDN architectures (designs): federated telco-CDNs and hybrid P2P-CDNs, using the dataset from two large Internet video providers. 
Wang et al. \cite{wang2017optimizing} exploits the use of Spark to largely analyze CDN log files to help content distributors make better decisions in how to improve QoS with minimum cost by utilizing multi-cloud CDN to serve users. Specifically, the authors present a multi-cloud architecture optimized for resource allocation and scheduling through big data analysis. They first analyze CDN log data on Spark to evaluate the QoS, then proposes two algorithms: one to perform a long-term resource allocation algorithm using minimum resources and another to handle burst demand when allocated resources are not enough. 

\section{Conclusions and Future Work}
\label{Sect:Conclusion}
In this paper, we analyze a real case study of a CDN vendor and have an insight into their CDN system. This work helps enterprises realize the potentials and issues of their system. The considered workload, which only contains log information in 7 days of one of the enterprise’s customers, is small and limited on the information of customer’s services so that we cannot propose more detailed analysis and better suggestions for the enterprise. In the future, we consider working with larger data with more various characteristics and details that can give us better insights into the enterprise’s system.

\vspace{-2pt}
\section*{Acknowledgment}
\vspace{-2pt}
This research was conducted within the project of Emulation of the color-based caching scheme in Telco-CDNs with Mininet using real data sponsored by TIS (IT Holding Group). 
\par
We acknowledge the support of time and facilities from Ho Chi Minh City University of Technology (HCMUT), VNU-HCM for this study.
\par We also acknowledge the support of time and facilities from High Performance Computing Laboratory, HCMUT.
\par
We would like to thank Thanh-Dang Diep from Ludwig Maximilian University of Munich for constructive criticism of the manuscript.

\bibliographystyle{IEEEtran}
\bibliography{IEEEfull,Refs}

\end{document}